# Spin control in heteromagnetic nanostructures

A.V. Scherbakov[1], A.V. Akimov[1], D.R. Yakovlev[2*], W. Ossau[3], L. Hansen[3], A. Waag[4], and L.W. Molenkamp[3]

[1]*A.F. Ioffe Physico-Technical Institute, Russian Academy of Sciences, 194021 St.Petersburg, Russia*

[2]*Experimental Physics 2, University of Dortmund, D-44227 Dortmund, Germany*

[3]*Experimental Physics 3, University of Würzburg, D-97074 Würzburg, Germany*

[4]*Institute of Semiconductor Technology, Braunschweig Technical University, 38106 Braunschweig, Germany*

[*] To whom correspondence should be addressed. E-mail: dmitri.yakovlev@physik.uni-dortmund.de

**The rapidly expanding research in *Spintronics*, the electronics utilizing the electron spin instead of its charge, is driven by the very interesting potential applications. The actual task is to develop principles for the spin manipulations in spintronic devices. In this *Report* we suggest and verify experimentally a concept of *heteromagnetic semiconductor structures*. It is based on spin diffusion between layers of the nanostructure with different magnetic properties and allows controlling the spin-switching rate for magnetic ions. A ten times increase of spin-lattice relaxation rate of magnetic Mn-ions is achieved in (Zn,Mn)Se/(Be,Mn)Te heteromagnetic structures with an inhomogeneous distribution of Mn-ions.**

Using the spins of free carriers instead of their charges for the operation of electronic devices is a promising concept for future electronics [1-3] and spintronics has rapidly become a well-established field of solid state physics [3]. A wide range of applications from storage of information to quantum computing and encrypting are envisaged for spintronic devices. The fabrication of new magnetic semiconductors suitable for room-temperature operation and the development of the principles for spin control of electrons and magnetic ions are active fields of research [3-8].

The interaction of spin systems with phonons, the vibrational waves in solids, is an important factor in spin dynamics [9]. Here, we introduce a concept enabling active control of the spin-phonon interaction of magnetic ions in semiconductor nanostructures. We verify this principle experimentally in (Zn,Mn)Se/(Be,Mn)Te heteromagnetic structures with an inhomogeneous distribution of magnetic Mn-ions. The suggested approach is based on the diffusion of spins between magnetic layers with different dynamical characteristics. This spin diffusion results in a strong reduction of the spin-lattice relaxation time and allows controlling the rate of spin-"switching" by structure parameters.

Control of the spin in semiconductor nanostructures requires methods for the generation of spin polarization, the manipulation of the spin state (e.g. moving the spin by a certain distance), as well as the destruction of spin polarization (i.e. a reset of the system into the initial state). Most of the magnetic structures that are used in spintronics contain magnetic ions such as manganese or chromium. Changing the polarization of the spin system implies a variation of the magnetization of the magnetic ions. For applications it is important to switch the spin polarization fast. This leads to the question of the time constants involved in spin-switching processes determined by the interaction of a spin with other elementary excitations. For instance, after switching on a magnetic field the spin system will be driven into a new equilibrium state characterized by a modified degree of spin polarization. However, the magnetization will not change instantaneously, because the transition time from the old to a new equilibrium state is governed by the interaction of the spin system with lattice phonons and is known as the spin-lattice relaxation time $\tau$ [9].

II-VI diluted magnetic semiconductors containing $Mn^{2+}$ magnetic ions are very reliable model systems for spintronics. They exhibit giant magneto-optical and magneto-transport effects [5,6]. The spin-lattice relaxation (SLR) in these materials occurs through a modulation of the exchange interaction between neighboring magnetic ions [10]. Thus the number of neighbors in the vicinity of the Mn ion is crucial for the SLR time. Characteristic SLR times vary from $10^{-8}$ to $10^{-3}$ s [11]. The SLR time depends on the lattice temperature and the applied magnetic field. However at fixed external conditions the Mn content is the main factor determining the spin relaxation [10]. The parameter, $x$, which describes Mn content, is the relative number of Mn ions substituting host cations (e.g. $Zn_{1-x}Mn_xSe$). It is often unfavorable to increase the Mn content higher than $x=0.01$ in the active layer of spintronic structures. Alloy fluctuations and associated defects induce carrier scattering and open nonradiative channels, compromising the transport and optical properties of the spintronic device.

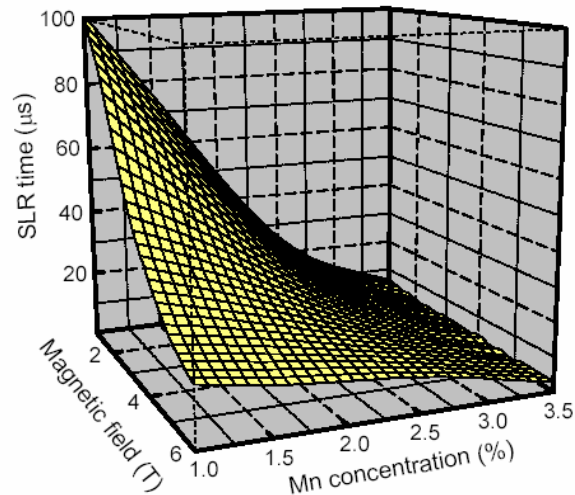

**Figure 1.** 3D-plot of the dependence of the spin lattice relaxation time of magnetic Mn-ions on magnetic field and Mn content for II-VI semiconductors. The plot is an interpolation of experimental data obtained for (Cd,Mn)Te at a temperature of 1.6 K (ref. 10).

Figure 1 shows the dependence of the SLR time $\tau$ on composition $x$ and magnetic field $B$ for (Cd,Mn)Te. Thus, in order to have a required value for $\tau$ in a certain spintronic device its design may differ. For instance, to have the spin-lattice relaxation as fast as possible, the Mn content should be rather high, $x>>0.01$. This in turn hampers the conducting and optical properties of the material. The dilemma is analogous to the optimization of the transport properties in semiconductors where the increase of carrier density obtained by excess doping is opposed by higher scattering rates of the carriers by impurity centers. Modulation-doped heterostructures, where the layer of dopants is separated from the active layer with free carriers, turned out to be the solution for this problem. Here, we suggest a magnetic counterpart of this concept; we introduce a *heteromagnetic semiconductor structure* containing different Mn contents in neighboring layers, which induces a strong decrease of the SLR time in the whole structure, while keeping $x$ in the active layer on a low level.

We use layer structures as shown in Fig.2. Sample A is a heteromagnetic structure with a non-homogeneous distribution of Mn ions. It consists of ten periods of 20 nm/10 nm $Zn_{0.99}Mn_{0.01}Se/Be_{0.93}Mn_{0.07}Te$ layers. Because of the band line-up, the layer with low Mn content ($Zn_{0.99}Mn_{0.01}Se$) acts as a quantum well for electrons. Sample B is a $Zn_{0.99}Mn_{0.01}Se/BeTe$ reference structure with no Mn in BeTe-based layers. Details of growth, band structure and optical properties of the samples may be found in Refs.12,13.

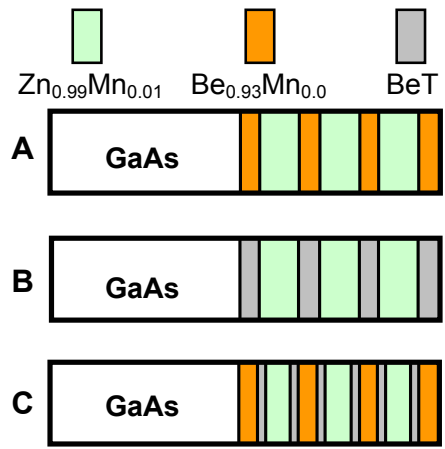

**Figure 2.** The II-VI semiconductor layered structures discussed in this paper: (A) a heteromagnetic structure where the Mn content is higher in orange layers; (B) the reference structure with nonmagnetic gray layers; (C) heteromagnetic layers with nonmagnetic spacers.

The temporal evolution of the magnetization, $M$, was measured by a method which combines spin excitation by a nonequilibrium phonon pulse and photoluminescence detection [10]. The type-II band alignment of (Zn,Mn)Se/(Be,Mn)Te heterostructures gives rise to spatially direct and indirect optical transitions, which are spectrally separated [12]. Monitoring of the two photoluminescence lines resulting from these transitions allows us to determine the magnetization in each particular layer. The direct luminescence probes the magnetization in the $Zn_{0.99}Mn_{0.01}Se$, while the indirect luminescence probes the $Be_{0.93}Mn_{0.07}Te$ layers in the sample A.

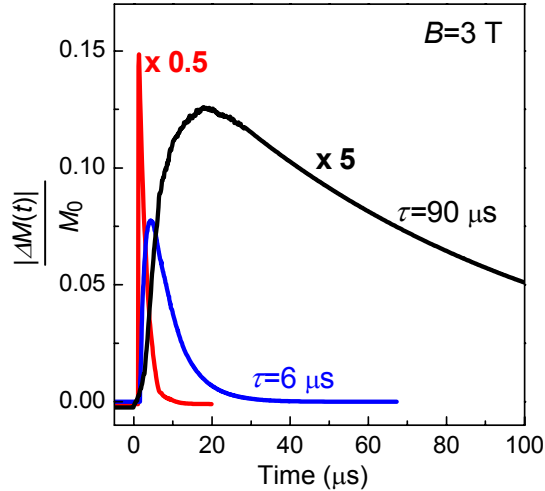

**Figure 3.** The time evolution of the relative changes in magnetization, $\Delta M(t)$, measured by monitoring the direct and indirect luminescence of samples A and B. The red (indirect) and blue (direct) curves were obtained from the heteromagnetic nanostructure A, and correspond to layers with high and low Mn content, respectively. The black curve is measured in the reference sample B, which contains nonmagnetic BeTe layers. The red and black curves are scaled for convenience.

Figure 3 shows the temporal changes of the magnetization caused by a phonon pulse in the Mn spin system which is pre-polarized by an external magnetic field of 3 T (bath temperature $T=1.6$ K). All signals exhibit a rise time which is the result of phonon-induced heating of the spin system accompanied by a decrease of the magnetization. The red line is the signal detected in the sample A on the indirect luminescence line and corresponds to changes of magnetization in $Be_{0.93}Mn_{0.07}Te$ layer. The signal has a decay of 1.5 μs which is about the duration of the phonon pulse. This is in agreement with the fact that the SLR time in $Be_{0.93}Mn_{0.07}Te$ is less than 1 μs. The blue and black lines show the time evolution of the magnetization in the $Zn_{0.99}Mn_{0.01}Se$ layers of samples A and B, respectively. The decay of the phonon-induced signal to its equilibrium value at bath temperature $T=1.6$ K and magnetic field $B=3$ T directly yields the SLR time of the Mn-ions. In the reference sample B $\tau=90$ μs, which is close to the value we obtained previously for bulk $Zn_{0.99}Mn_{0.01}Se$ [14]. The presence of the magnetic $Be_{0.93}Mn_{0.07}Te$ layers in sample A leads to a drastic decrease of the SLR time by more than one order of magnitude to $\tau=6$ µs. This result clearly demonstrates that strong acceleration of the spin-lattice relaxation does indeed occur in heteromagnetic semiconductor structures.

The amplitude of the phonon-induced signal is three times higher in the heteromagnetic structure A than in the reference sample B (cf. the amplitudes for the blue and black lines in Fig.3, taking into account the scaling factor). This means that for a similar excitation (in our case the same phonon pulse) the magnetization changes are larger in the heteromagnetic structure. This is useful for spintronic devices where not only fast but also large changes in spin polarization are required.

Next, we turn to the physical mechanism of the observed effect. The spin-spin interaction between neighbouring $Mn^{2+}$ ions can give rise to a spin-flip transition [9,11]. This is a transition where the excited spin of a Mn ion relaxes to the ground state while the neighbouring spin becomes excited. As a result an excited spin "jumps" from one Mn ion to another or, in other words, the spin excitation diffuses. In layers with low $x$, such a spin diffusion over distances of about 10 nm occurs much faster than the SLR process [15]. This means that in heteromagnetic nanostructures, an excited spin in the layer with low $x$ diffuses rapidly to the layer with high $x$ value, where the SLR is fast. In other words, the presence of the layer with high $x$ accelerates the overall relaxation of the spin polarization in the structure.

The observed effect may be exploited in a spintronic device containing layers with low and high $x$, which we refer to as the "active" (for electron transport) and "spin dump" layers, respectively. Figure 4 schematically shows the role that spin diffusion plays in such an heteromagnetic nanostructure. In a homogeneous magnetic structure, or in a structure with nonmagnetic BeTe layers, spin-phonon transitions only occur in the active layers (Fig. 4B). A phonon is absorbed during the heating process and emitted when a reverse transition occurs. Phonon absorption in a heteromagnetic structure occurs mainly in the "spin dump" layer (left part of Fig. 4A), because the spin phonon transition is more probable there due to high Mn content. Subsequently, the excited spin diffuses to the active layer. The green horizontal arrows in Fig.4A demonstrate this diffusion. Cooling occurs in the reverse order (right part of Fig. 4A. Thus the spin polarization in the active layer changes by a larger value during the excitation and relaxes faster in a heteromagnetic nanostructure than in a homogeneous structure.

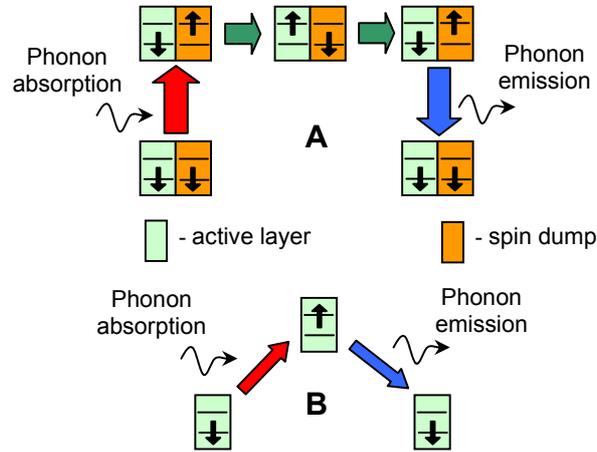

**Figure 4.** A diagram demonstrating the phonon absorption and emission processes in (A) a heteromagnetic nanostructure and (B) a homogeneous magnetic structure. The red and blue arrows indicate spin-phonon transitions. Green arrows in (A) show the spatial spin diffusion from the "spin dump" to the "active" layer (spin excitation) and back (relaxation of the excited spin).

The suggested device concept provides an effective and simple way to control the SLR times in spintronic devices. An active layer with a specific required Mn content may be grown in contact with another, "spin dump", layer where the Mn content is high. The presence of the latter does not influence the carrier properties in the active layer but significantly reduces the SLR time. A more precise control may be realized in heteromagnetic structures with nonmagnetic spacer layers (sample C in Fig. 2). In this case the spin transfer time between the magnetic layers depends on the spacer thickness and thus the spin relaxation time in the active layer may be adjusted from very short (no spacer) to quite long (wide spacers or nonmagnetic layers).

In our experiments we have chosen structural parameters and experimental conditions which enable us to measure SLR times using a relatively slow experimental method. In real spintronic devices the relaxation processes may be much faster, depending on the design of the heteromagnetic nanostructure and its working conditions. Finally, we note that the principle of operation of the heteromagnetic nanostructures proposed here is not limited to low temperatures.